\documentclass[sigconf]{acmart}

\usepackage{float}
\usepackage{amsmath,amsfonts,mathtools}
\usepackage{algpseudocode}
\usepackage{algorithm}
\usepackage{array,multirow,graphicx}
\usepackage{textcomp}
\usepackage{subfig}
\usepackage{booktabs}
\usepackage{enumitem}
\usepackage{threeparttable}
\usepackage{todonotes}
\usepackage[multiple]{footmisc}
\usepackage{balance}
\usepackage{xcolor,colortbl,soul}
\usepackage{hyperref}
\usepackage{makecell}
\usepackage{afterpage}

\AtBeginDocument{%
  }

\definecolor{Gray}{gray}{0.85}
\newcolumntype{a}{>{\columncolor{Gray}}c}

\DeclareMathAlphabet{\altmathcal}{OMS}{cmsy}{m}{n}

\sethlcolor{lightgray}

\setcopyright{acmcopyright}
\copyrightyear{2024}
\acmYear{2024}
\setcopyright{rightsretained}
\acmConference[ICPE '24]{Proceedings of the 15th ACM/SPEC International Conference on Performance Engineering}{May 7--11, 2024}{London, United Kingdom}
\acmBooktitle{Proceedings of the 15th ACM/SPEC International Conference on Performance Engineering (ICPE '24), May 7--11, 2024, London, United Kingdom}\acmDOI{10.1145/3629526.3645048}
\acmISBN{979-8-4007-0444-4/24/05}

\begin{document}

\title{Demeter: Resource-Efficient Distributed Stream Processing under Dynamic Loads with Multi-Configuration Optimization}

%\author{Anonymous Author(s)}

\author{Morgan K. Geldenhuys}
\affiliation{%
   \institution{Technische Universität Berlin}
   \city{Berlin}
   \country{Germany}
 }
 \email{morgan.geldenhuys@tu-berlin.de}

 \author{Dominik Scheinert}
 \affiliation{%
   \institution{Technische Universität Berlin}
   \city{Berlin}
   \country{Germany}
 }
 \email{dominik.scheinert@tu-berlin.de}

 \author{Odej Kao}
 \affiliation{%
   \institution{Technische Universität Berlin}
   \city{Berlin}
   \country{Germany}
 }
 \email{odej.kao@tu-berlin.de}

 \author{Lauritz Thamsen}
 \affiliation{%
   \institution{University of Glasgow}
   \city{Glasgow}
   \country{United Kingdom}
 }
 \email{lauritz.thamsen@glasgow.ac.uk}

\begin{abstract}
Distributed Stream Processing (DSP) focuses on the near real-time processing of large streams of unbounded data. 
To increase processing capacities, DSP systems are able to dynamically scale across a cluster of commodity nodes, ensuring a good Quality of Service despite variable workloads.
However, selecting scaleout configurations which maximize resource utilization remains a challenge. 
This is especially true in environments where workloads change over time and node failures are all but inevitable.
Furthermore, configuration parameters such as memory allocation and checkpointing intervals impact performance and resource usage as well.
Sub-optimal configurations easily lead to high operational costs, poor performance, or unacceptable loss of service. 

In this paper, we present Demeter, a method for dynamically optimizing key DSP system configuration parameters for resource efficiency. 
Demeter uses Time Series Forecasting to predict future workloads and Multi-Objective Bayesian Optimization to model runtime behaviors in relation to parameter settings and workload rates. 
Together, these techniques allow us to determine whether or not enough is known about the predicted workload rate to proactively initiate short-lived parallel profiling runs for data gathering.
Once trained, the models guide the adjustment of multiple, potentially dependent system configuration parameters ensuring optimized performance and resource usage in response to changing workload rates.
Our experiments on a commodity cluster using Apache Flink demonstrate that Demeter significantly improves the operational efficiency of long-running benchmark jobs.

\end{abstract}




\maketitle

\section{Introduction}

Distributed Stream Processing (DSP) is a paradigm in the big data domain that focuses on processing large streams of unbounded data in near real-time. 
As streaming workloads are typically dynamic in nature, DSP systems such as Apache Flink~\cite{carbone2015apache}, Spark Streaming~\cite{Zaharia2016ApacheS}, and Storm~\cite{Toshniwal2014Stormtwitter} are designed to scale horizontally, distributing the load across multiple nodes within a cluster. 
When configured correctly, this enables high throughput, low latency, and fault tolerance in cloud-based environments.
This is essential in areas such as real-time analytics, IoT data processing, click stream analysis, network monitoring, and more, where data needs to be analyzed on-the-fly continuously~\cite{Nasiri2018ASO, IAM+19, NNG19}. 
In these areas, ensuring minimal latencies is vital as the value of results is often greatest at the time of data arrival.
Therefore, by maintaining the fastest response times, the maximum value of these results is captured, enabling timely decision-making.
Likewise, in the event of failures, the ability to both recover quickly and maintain the consistency of results is important for preserving system reliability.

However, the manner in which these systems are configured has a significant impact on the Quality of Service (QoS) they are able to deliver as well as the resources they consume. 
Given the dynamic nature of cluster environments and streaming workloads, ensuring near-optimal configurations becomes inherently challenging.
Relying on static configurations can lead to over-provisioning, wasting energy and resources, or under-provisioning.
The manual fine-tuning of configuration parameters for individual streaming jobs is likewise impractical, involving considerable trial and error, and necessitates continuous adjustments to align with the ever-changing workload.
As a result, configurations would quickly become outdated, leading to a diminished QoS and significantly higher operational costs.
It is evident that adaptive configuration optimization strategies are essential, as they can dynamically respond to changing workloads, ensuring both reliability and efficiency.

A number of methods have been proposed that optimize the configuration of DSP jobs adaptively. 
The underlying principle is to identify a configuration parameter that, when adjusted, can improve the performance of the DSP job.
For instance, some methods focus on optimizing the scaleout~\cite{flink2023reactive, databricks2023elasticSpark, GSH+14, FAG+17, PETROV2018109, HuKZ19, Kalavri2018ThreeSI, Geldenhuys2022PhoebeQD}, while others concentrate on adjusting the checkpoint interval~\cite{ Y74,D03,D06,JHK20}.
They can broadly be categorized as either reactive or proactive. 
Reactive strategies~\cite{GSH+14,FAG+17} typically rely on threshold monitoring, adjusting the parameter when the established upper or lower bounds are violated. 
However, these approaches are imprecise and prone to initiating re-configurations in response to transient conditions such as load spikes or failures. 
This can lead to unnecessary changes that disrupt the service.
Alternatively, proactive strategies~\cite{flink2023reactive, databricks2023elasticSpark, PETROV2018109,HuKZ19,Kalavri2018ThreeSI,Geldenhuys2022PhoebeQD} often employ modeling techniques, using historical data to predict a near-optimal configuration setting for the workload.
Nonetheless, these methods are not without their challenges, especially when historical data is limited, complicating accurate predictions in dynamic environments.

Although these methods have shown improvements in certain aspects of system performance, the focus on a single parameter does not fully capture the complex interdependencies among various key configuration parameters that are common in real-world DSP systems.
Take, for example, the concepts of scaleout and local parallelism.
In a DSP system, the scaleout for each job is reflected by the total number of processing slots, and the ratio of processing slots to each worker node determines the level of local parallelism.
Generally, previous methods have kept this ratio at one-to-one, resulting in the total processing slots equalling the number of workers. 
They do not explore how local parallelism might boost efficiency beyond identifying the optimal scaleout for a particular workload rate. 
Despite these interdependencies, the simultaneous optimization of multiple configuration parameters for enhancing resource efficiency in DSP systems remains largely unexplored.

In this paper, we introduce \textit{Demeter}, a novel approach intended for cloud-based environments that dynamically optimizes multiple key configuration parameters for targeted DSP jobs. 
We focus on the following critical parameters:

\begin{itemize}%[leftmargin=*]

    \item \textbf{No. of Workers:} Workers are responsible for orchestrating and supervising the execution of processing tasks. Their configuration affects the overall capacity of the system to handle parallel tasks and its resilience against failures.
    
    \item \textbf{CPU Cores:} These refer to the computational resources assigned to each worker.
    Workers receive an equal share of computation resources, ensuring homogeneous processing potential throughout the system.

    \item \textbf{Memory Allocation:} Memory is important for buffering incoming data streams, storing intermediate results, and facilitating various in-memory operations. 
    As with the CPU cores, each worker is assigned an equal share of memory.

    \item \textbf{Processing Slots:} A worker can manage multiple processing slots, determining the number of concurrent processing tasks it can oversee. These slots utilize CPU and memory resources from the collective resource pool assigned to each worker.

    \item \textbf{Checkpoint Interval:} This involves periodic snapshots of the system's current state as a part of its fault-tolerance mechanism. 
    The checkpoint frequency is important; while more frequent snapshots can facilitate faster recovery from failures, they also introduce overheads.
    
\end{itemize}

Demeter aims to find the best combination of settings for the aforementioned parameters, ensuring both near-optimal performance and resource efficiency in response to changing workloads.
To achieve this, Demeter makes use of two statistical modeling techniques: Time Series Forecasting (TSF) for predicting workload rates, and Multi-Objective Bayesian Optimization (MOBO) to model runtime behaviors in relation to parameter settings and workload rates.
From a high-level perspective, Demeter operates through two independent processes: \emph{profiling} and \emph{optimization}. 
Both processes execute iteratively and begin with a prediction of the upcoming workload rate.
In profiling, when the MOBO models lack sufficient data for confidently selecting a near-optimal configuration at the predicted rate, Demeter initiates short-lived parallel profiling runs to gather data to enhance the models. 
On the other hand, during optimization, once the models are sufficiently informed about the predicted workload rates, Demeter selects appropriate configurations.
The objective is to minimize resource usage such that processing latencies and recovery times remain within the bounds of a runtime-derived latency constraint and a user-defined recovery time constraint.
With continued iterations, Demeter's understanding of how configuration impacts on performance at specific workload rates increases along with resource efficiency.
Over time, as the models become more accurate, the need for profiling decreases, further decreasing overheads.
We implement Demeter prototypically with Apache Flink and conduct experiments on a commodity cluster with established benchmark jobs to demonstrate its usefulness in comparison to two state-of-the-art methods.

\begin{figure*}[!h]
    \centering
    \includegraphics[width=1.0\textwidth]{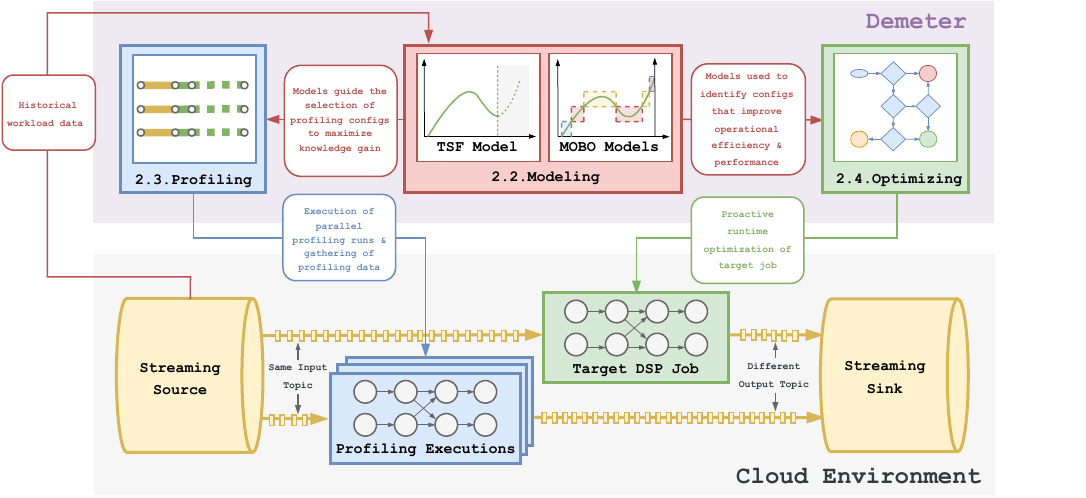}
    \caption{High-level representation of Demeter, illustrating the interplay between internal processes and external systems.}
    \label{fig:overview}
\end{figure*}

\section{Approach}

In this section, we provide a detailed description of Demeter, explaining the general idea and its processes.

\subsection{General Idea}

Demeter's goal is the runtime optimization of a target DSP job, ensuring near-optimal performance and resource efficiency across dynamic workloads.
We measure resource efficiency in terms of \textit{resource usage} ($U$) and performance in terms of the \textit{average end-to-end latency} ($L_{avg}$) and \textit{recovery time} ($R$).
The process is guided by two constraints: an \textit{average end-to-end latency constraint} ($LC$) which is determined at runtime based on observed latencies; and a user-defined \textit{recovery time constraint} ($RC$).
Therefore, the objective is to minimize $U$ while ensuring $L_{avg}$ and $R$ are always kept below $LC$ and $RC$, respectively.
The decision variables consist of a set of key configuration parameters: number of workers, CPU cores, memory allocation, processing slots, and checkpoint interval.

To achieve its goal, Demeter employs a proactive modeling strategy, gathering performance data through short-lived parallel profiling runs of identical jobs with varying configuration sets.
This strategy is enabled by using two fundamental statistical modeling techniques: Time Series Forecasting (TSF) to predict upcoming workload rates and Multi-Objective Bayesian Optimization (MOBO) to model runtime behaviors in relation to parameter settings and workload rates.
TSF grants insight into upcoming workloads, facilitating informed decision-making.
This is important for preventing performance from degrading beyond a critical point before mitigating actions can be taken and ensures the longevity of configuration changes, thereby reducing the frequency of restarts.
When working with exactly-once processing guarantees, restarts are expensive and introduce interruptions to the service.
MOBO complements this by not only providing a means of simultaneously modeling multiple possibly competing criteria, but also a mechanism for guiding the exploration of the configuration search space.
Together, TSF and MOBO establish a comprehensive approach for well-informed multi-configuration optimization under dynamic workloads.

Demeter is designed as a standalone client, interfacing with systems within a cloud-based environment.
Our methodology is built on three foundational processes: Modeling, Profiling, and Optimizing. 
Next, we present an overview of Demeter's approach, accompanied by a graphical representation in Fig.~\ref{fig:overview}.
For any target job, a \textit{maximum configuration} (\( C_{\text{max}} \)) is defined, where parameters are set to allocate a large amount of resources, thereby guaranteeing consistent high performance in terms of \( L_{\text{avg}} \) and \( R \) across any reasonable workload\footnote{Here, 'reasonable' refers to workloads within an upper bound determined by expert knowledge.}. 
After initiating the target job with \( C_{\text{max}} \), two iterative processes begin executing asynchronously.
The first process focuses on profiling. 
Based on the predicted workload rate, the need for profiling is evaluated through a series of MOBO models, with each model dedicated to a particular configuration parameter.
When required, they suggest configurations that maximize information gain, which are then applied in parallel profiling runs. 
This approach enables efficient exploration of the large configuration space. 
After these runs, data is collected to update the models. 
If profiling is deemed unnecessary, it is skipped.
The second process focuses on optimization. 
It uses the TSF prediction and the MOBO models to check whether sufficient information is available for job optimization at that workload rate. 
If a more efficient configuration is found, a reconfiguration is initiated. 
If no better configuration is available according to our models but the existing setup can manage the upcoming workload, it remains unchanged. 
In situations where the current configuration is inadequate, or if there is insufficient information, such as encountering a new workload rate, the system reverts to the \( C_{\text{max}} \) configuration, unless it is already in use.

\subsection{Modeling}
\label{sec:approach_modeling}

In this section, we provide a description of the modeling techniques used in our approach. 
As previously stated, Demeter follows a proactive strategy, using predictions of future workload rates to inform and guide the profiling and optimization processes.
To achieve this, we firstly use a multistep-ahead TSF model, trained on historical data. 
This model not only provides insights into the expected workload rate at a specific time horizon but also reveals the behavior of the rate over time leading up to this point.
Understanding both the expected future workload rate and its behavior over time is important for our method's effectiveness.
Whenever a forecast is generated, it is partitioned into separate averaging bins and the bin with the highest average value is calculated and selected.
The value of this bin will be used for all subsequent profiling and optimization processes.
This ensures that if an increase in workload is anticipated, the system will select the furthest bin for analysis, guiding the profiling and optimization efforts to address higher workload rates expected in the future, rather than focusing on the current point in time.
Conversely, if the workload is anticipated to decrease, the model selects the nearest bin, focusing the profiling and optimization efforts on lower workload rates that are expected to occur closer to the current point in time.
This is done to ensure that reconfigurations are valid at least up until the forecast horizon, thereby increasing their longevity.
For our method, we use an online ARIMA model for workload predictions.

Addressing the challenge of optimizing multiple parameters, our approach shifts focus towards exploring a discrete search space, represented as $\mathbf{x}$. 
This space consists of the Cartesian product of all relevant parameters, each with its discrete set of potential values. 
Initially, we have no prior knowledge of this space. 
In anticipation of future streaming workloads, our primary goal is to identify an optimal configuration within $\mathbf{x}$. 
Such a configuration should not only comply with predefined constraints, including \(LC\) and \(RC\), but also demonstrate resource efficiency, in terms of CPU and memory utilization.
To address this challenge, we adopt MOBO, where each objective and constraint is represented using an individual Gaussian process. 
The optimization procedure aims to maximize the sum of their marginal log likelihoods. 
In this context, we weight the expected improvement of one or more objectives by the probability of feasibility, considering all the modeled constraints.
Ultimately, the MOBO model is proficient in identifying near-optimal configurations for scenarios characterized by low workload variance.
However, its performance diminishes when dealing with streaming workloads that present a wider range of variance, highlighting the need for a more adaptable optimization strategy.

\begin{figure}
    \centering
    \includegraphics[width=\columnwidth]{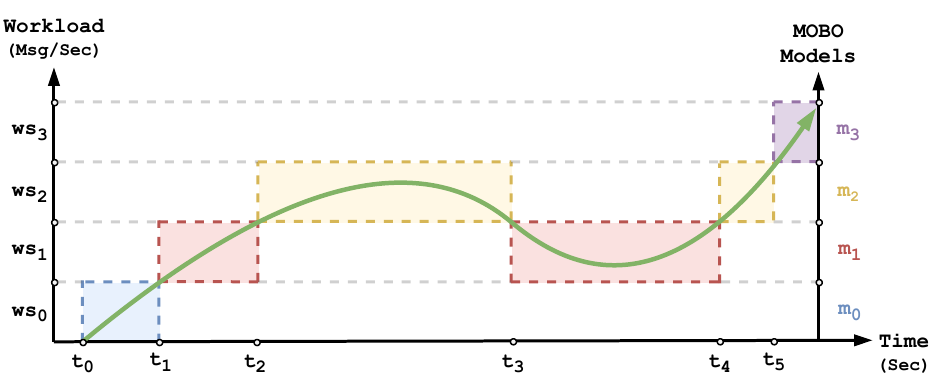}
    \caption{Model generation and selection based on observed workload rates, categorized by specific workload segments.}
    \label{fig:modeling}
\end{figure}

To manage workload variability, we apply the concept of Rank-Weighted Gaussian Process Ensembles (RGPE), a method proven effective in similar scenarios~\cite{ZhangWCJT0Z021,ScheinertWWTWK2023}. 
All collected observations, comprising various configurations and their corresponding performance metrics, are organized into $K$ segments based on workload rate, denoted as $WS=\{ws_i\}_{i=0}^{K}$.
The size of each workload segment is defined by the \textit{segment size} ($SS$) hyper-parameter, and segments are added dynamically as they are detected.
We train a set of MOBO models $m_i$ for each segment $ws_i$ using the included observations $D_i$.
Referring to Fig.~\ref{fig:modeling}, we present an example of a variable workload over time. 
As Demeter identifies new workload segments \(ws_{0}\), \(ws_{1}\), \(ws_{2}\), and \(ws_{3}\) at time instances \(t_{0}\), \(t_{1}\), \(t_{2}\), and \(t_{5}\), it correspondingly creates MOBO models \(m_{0}\), \(m_{1}\), \(m_{2}\), and \(m_{3}\) to model the configuration parameters for these segments.
Following this, whenever the workload prediction falls within a segment’s range, the system utilizes the associated MOBO models for both profiling and optimization purposes.
These models then contribute to approximating the target MOBO model $m_{tar}$ for the anticipated streaming workload in an ensemble manner:
\begin{equation*}
m_{tar}(\mathbf{x}|D_{tar}) \sim \altmathcal{N}\bigg( \sum_{ws_i \in WS} a_i \mu_i(\mathbf{x}), \sum_{ws_i \in WS} a_i^2 \sigma_i^2(\mathbf{x}) \bigg)
\end{equation*}
In this formulation, $\mu_i$ and $\sigma_i^2$ are the mean and variance parameters of the BO model $m_i$, tailored for segment $ws_i$, while $a_i$ represents the corresponding weight within the ensemble, defined by a ranking loss specific to RGPE. 
This approach offers several benefits, including the ability to leverage previously trained models, effectively addressing the cold-start issue by utilizing existing knowledge, even if partial, of the configuration search space. 
Concurrently, these \emph{support} models play a crucial role in pinpointing the optimum of the desired function.
Consequently, we expedite the identification of promising configurations by utilizing previously profiled configurations across varied workload rates. 
So, our outlined approach, combined with RGPE, allows us to leverage existing knowledge for more informed configuration recommendations.
For simplicity, in this paper, the RGPE ensemble model, target MOBO models, and base MOBO models for a specific segment are collectively referred to as the MOBO models for that segment.

\subsection{Profiling}
\label{sec:approach_profiling}

The profiling process evaluates the need for profiling, selects fitting configurations, and supervises execution. 
Operating within a time-delay loop, the profiling algorithm begins each iteration by consulting the TSF model for a workload prediction. 
This prediction then guides the selection of configurations by referencing the MOBO models corresponding to the specific workload segment.
For generating profiles, the method focuses on configurations with the highest expected hyper-volume improvement, essentially those with the best knowledge acquisition value. 
The number of configurations chosen for profiling aligns with the resources assigned for this task within the cluster. 
Within each workload segments, an annealing factor adjusts this number, leading to a reduced extent of exploration as more knowledge about that segment is accumulated.

Additionally, the system uses domain knowledge from past configurations to make strategic profiling decisions. 
For example, if during a previous re-configuration, a selection was found to be unsuitable for a workload rate similar to the prediction, and the system had reverted back to the \( C_{\text{max}} \), it would then give preference to profiling configurations that have greater resources than the previously unsuitable one.
On the other hand, should historical data show an earlier decision to downscale, the system will prioritize configurations that use fewer resources for the projected workload. 
This approach allows the system to streamline its selection process, capitalizing on historical data and specialized expertise.

\begin{figure}
    \centering
    \includegraphics[width=\columnwidth]{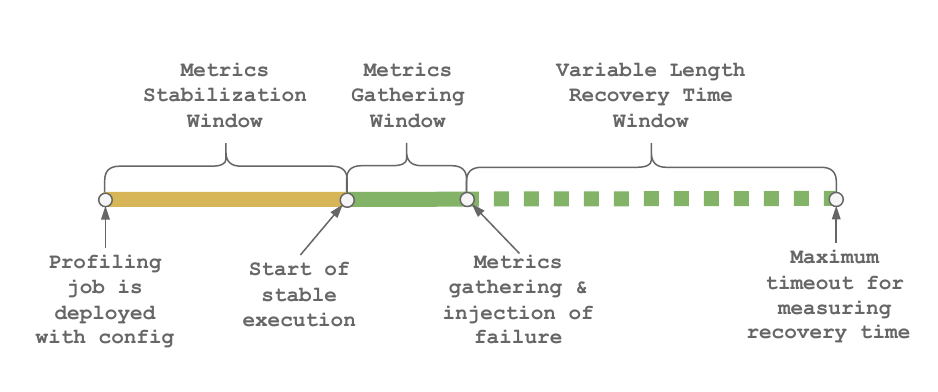}
    \caption{Profiling lifecycle from deployment to recovery to recovery time measurement.}
    \label{fig:profiling}
\end{figure}

Upon determining suitable profiles for the predicted workload, jobs are deployed in parallel with their own unique configurations. 
A graphical representation of the profiling process can be seen in Fig.~\ref{fig:profiling}.
Each profile is set to read from the same data source as the target job, yet directs its output to a separate temporary sink. 
Once deployed, a stabilization phase is allowed for metrics to achieve equilibrium. 
After stabilization, a fixed duration of standard execution is observed, after which the \(L_{\text{avg}}\) are computed. 
Next, timeout failures are injected into the profiling jobs to measure how long each job takes to recover. 
Demeter will then continuously re-evaluate the status of the jobs, ensuring they either attain a full recovery or exceed a designated maximum timeout constraint. 
In order to measure the recovery time, Demeter monitors specific metrics related to the recovery process:

\begin{itemize}%[leftmargin=*]

    \item \textbf{Input Throughput:} This metric indicates the total number of events consumed by the source operators of the DSP job every second. It provides a measure of the system's capacity to handle incoming data.

    \item \textbf{Average Consumer Lag:} This metric represents the accumulated events in the messaging queue, still awaiting consumption by the source operators of the DSP job. It gives an indication of any potential backlog.

\end{itemize}

These metrics are used to train an anomaly detection algorithm on positive executions, i.e. let the function $s : X \rightarrow X$ perfectly represent the metrics data stream such that for any given data point $x \in X$ the prediction is always $s(x) = x$.
Given that the majority of data collected within the standard execution period is expected to be normal, this approach allows the algorithm to recognize deviations from the norm. 
Should these deviations surpass a predefined threshold, derived from past errors, the system is deemed to be in an anomalous state. 
The length of time spent in this anomalous state therefore is equivalent to the recovery time. 
To implement this, we utilize an online ARIMA method as proposed in~\cite{SSG+18}.

Importantly, in this context the recovery time encompasses more than just the period during which the system is in an inconsistent state before processing resumes.
For systems using checkpoint and rollback recovery strategies, processing restarts from a previously saved offset.
It then works to catch up to the latest offset, even as new events keep arriving. 
We aim to measure the entire duration – from the moment the failure begins until processing has once again caught up to processing events at the latest offset. 
This provides a more accurate measure of system availability for steam processing.

Upon the completion of profiling, the associated jobs are terminated, and metrics relating to workload, throughput, and latency are subsequently used to update the model.

\subsection{Optimizing}
\label{sec:approach_optimizing}

\begin{figure}
    \centering
    \includegraphics[width=\columnwidth]{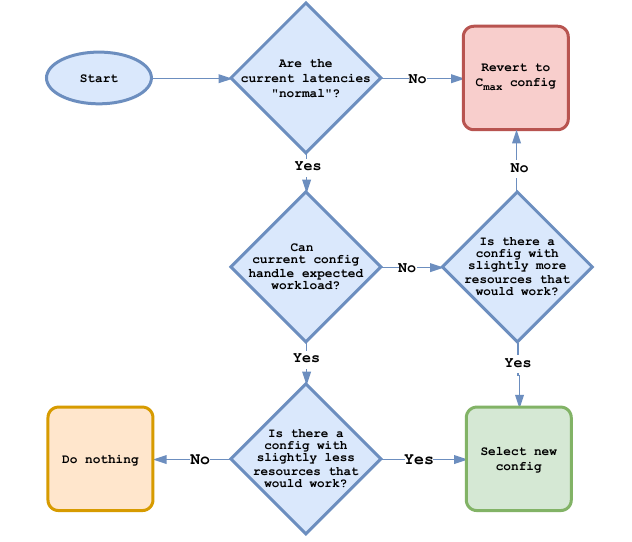}
    \caption{Flow chart depicting the optimization algorithm of the Optimizer component.}
    \label{fig:optimizing}
\end{figure}

The optimization process is tasked with tuning the configuration parameters of the target job in relation to changing workloads. 
Central to this process is the optimization algorithm, which operates within a time-delayed loop to periodically assess the state of the target job. 
Intervals between evaluations are essential, particularly following events like re-configurations or failures that both require restarts. 
They ensure that metrics stabilize between evaluations and increase the longevity of changes.
A simplified overview of this algorithm can be seen in \autoref{fig:optimizing}.
After each interval, metrics from the target job are collected and the current \(L_{\text{avg}}\) is evaluated. 
Unlike our other constraints and objectives, determining whether the \(L_{\text{avg}}\) aligns with typical expectations presents a challenge, primarily because 'normal' latencies vary significantly, changing from one job and environment to another.
To address this, we utilize the MOBO models, which are trained on the specific current workload rate segment, in combination with a clustering technique to establish a benchmark for near-optimal latencies.

Firstly, we need to identify two clusters among the latencies observed so far: those considered normal and those considered abnormal, and hence invalid.
If a configuration resulted in the job being able to keep up with the current workload, its latencies would stabilize around the shortest possible lengths. 
Thus, the cluster with the smallest centroid represents configurations that yield near-optimal latencies.
In order to do this, we start by normalizing the values according to the first percentile. 
Then, we apply a transformation to the values so that they fall within the $[0,1]$ interval. 
Values smaller than 0.5 are considered normal, while those equal to or greater than 0.5 are considered abnormal.
This technique allows us to define the latency constraint \(LC\).
If the current \(L_{\text{avg}}\) falls outside the range of \(LC\), the job is assumed to be unstable and reverts to the \(C_{\text{max}}\) configuration, ensuring a rapid return to near-optimal operation.

Alternatively, if \(L_{\text{avg}}\) falls within acceptable limits, there exists an opportunity to further improve resource efficiency.
In both cases, we transform all latency values according to the observed clusters, establishing a notion of normal and invalid latencies.
The next step involves retrieving a workload prediction from the TSF model. 
We now aim to identify the near-optimal configuration that aligns most closely with the predicted workload. 
Using this workload prediction, the algorithm references the MOBO models for the corresponding workload rate segment. 
We then use the models to retrieve predictions \(L_{\text{avg}}\) and \(R\) for a list of possible configurations.

We first filter out configurations with invalid \(L_{\text{avg}}\) as well as those where the \( R \) is greater than the \(RC\), ensuring we retain only valid configurations that do not violate either constraint.
Next, we sort the list of valid configurations based on their resource usage, where the smallest is predicted to result in the highest operation efficiency. 
At this point, we introduce a \textit{safety buffer} (\(SB\)) hyper-parameter. 
Instead of directly selecting the most efficient configuration, \(SB\) is applied to scale our selection, effectively moving us up the sorted list by a certain percentile. 
For instance, if \(SB\) is set to 30\%, we would skip the bottom 30\% of configurations and choose the one at the 30th percentile mark, or with at least a 30\% increase in resource usage. 
This provides a margin of safety, ensuring that our selected configuration is not too close to the lower bound of the system's requirements. 
Furthermore, this buffer approach allows us to account for and negate intermittent fluctuations in the cluster, which could otherwise trigger unwarranted re-configurations, optimizing for both efficiency and system stability.

Following our selection process informed by \(SB\), the need for re-configuration is evaluated against the \textit{efficiency threshold} (\(ET\)) hyper-parameter. 
Reconfiguration is triggered only if the expected improvement in resource efficiency exceeds \(ET\). 
For example, setting \(ET\) at 5\% means reconfiguration is only done when the resource saving is equal to or greater than 5\%. 
Therefore, if the proposed configuration fails to meet this threshold, the current configuration remains unchanged, and the current iteration concludes. 
Conversely, if the threshold is exceeded, signifying a potential reduction in resource usage, reconfiguration is initiated. 
In instances where a predicted configuration is not available, potentially due to a lack of sufficient observations for modeling, a re-configuration to the \(C_{\text{max}}\) configuration is prompted, provided it is not already in place.

\section{Evaluation}

In this section, we assess the effectiveness of Demeter. 
We detail our experimental cluster setup, outline the methods evaluated, and specify how they were configured. 
Two distinct experiments are presented, followed by a comparative analysis. 
All materials related to our prototype, datasets, and tools are available in our repository\footnote{\url{https://github.com/dos-group/demeter}}

\subsection{Experimental Setup}

Our experimental setup was based on a co-located 5-node Kubernetes~\cite{VPK+15} and HDFS~\cite{SKR+10} cluster with all servers interconnected by a single switch. 
We developed and implemented a prototype to work with Apache Flink.
Additionally, we use the Flink Kubernetes Operator\footnote{\url{https://nightlies.apache.org/flink/flink-kubernetes-operator-docs-release-1.6}, Accessed: March 2024} to automate deployments and upgrades of streaming jobs.
We configured an Apache Kafka~\cite{KNR11} cluster, serving as both the sources and sinks for the streaming jobs, with 24 partitions and a replication factor of 3.
All sources and sinks of the experimental processing pipelines were configured to use \textit{exactly-once} processing thereby guaranteeing the consistency of results~\footnote{\url{https://flink.apache.org/2018/02/28/an-overview-of-end-to-end-exactly-once-processing-in-apache-flink-with-apache-kafka-too/}, Accessed: March 2024}.
For all experiments and methods, a maximum parallelism of 24 was set, meaning that the number of processing slots, or task managers, could not exceed 24 at any given time.
Additionally, we set a 20s timeout interval for Flink task managers. 
For end-to-end latencies, measurements were taken over a 1-minute averaging window, with a focus on the 95th percentile to minimize the impact of outliers during periods of stable operation. 
Each experiment was conducted three times with the median result being selected for further analysis and discussion.
Chaos Mesh\footnote{\url{https://chaos-mesh.org}, Accessed: March 2024} was used for injecting failures into the Kubernetes pods. 
During each experiment, Chaos Mesh injected 23 timeout failures at regular 45-minute intervals ensuring a uniform distribution of failures across a broad range of workload rates.
Prometheus\footnote{\url{https://prometheus.io}, Accessed: March 2024} was used for metrics collection.
Cluster node specifications and software versions are summarized in~\autoref{tbl:clusterspecs}.

\begin{table}[H]
\centering
\caption{Cluster Node Specifications}
\begin{tabular}[t]{rp{0.65\linewidth}}
    \toprule
    \textbf{Resource}&\textbf{Details}\\
    \midrule
    OS&Ubuntu 20.04.1\\
    CPU&AMD EPYC 7282 16-Core Processor, 32 cores, 2.8 GHz\\
    Memory&128 GB RAM\\
    Storage&2TB RAID0 (2x1TB SSD, software RAID)\\
    Network&10 GBit Ethernet NIC\\
    Software&Java v11, Flink v1.17, Flink Operator v1.6, Kafka v3.4, Docker v19.3, Kubernetes v1.26, HDFS v2.8, Redis v5.0, Prometheus v2.25, Chaos Mesh v2.1, pmdarima v2.0.4, BoTorch v0.6.0
    \\
    \bottomrule
\end{tabular}
\label{tbl:clusterspecs}
\end{table}
\setlength{\textfloatsep}{0.1cm}

\subsection{Demeter Setup}

In configuring Demeter, the \( C_{\text{max}} \) configuration was allocated a maximum scaleout of 24, dedicating a full CPU core and 4096 megabytes of memory to each taskmanager, as well as a single processing slot.
The \textit{segment size} (\(SS\)), \textit{safety buffer} (\(SB\)), and \textit{efficiency threshold} (\(ET\)) hyper-parameters were set to 10.000, 30\%, and 5\%, respectively. 
The configuration space for profiling was defined by setting lower and upper bounds on each parameter, yielding 2592 distinct parameter combinations for each workload segment, as shown in~\autoref{tbl:config_parameters}. 
We utilize BoTorch~\cite{balandat2020botorch} for MOBO modeling as part of our prototype.

\begin{table}[H]
\centering
\caption{Configuration Parameter Search Space}
\begin{tabular}[t]{llll}
    \toprule
    \textbf{Parameter} & \textbf{Min} & \textbf{Max} & \textbf{Step} \\
    \midrule
    Workers & 4 & 24 & 4 \\
    CPU Cores & 1 & 3 & 1 \\
    Memory Allocation (mb) & 1024 & 4096 & 1024 \\
    Processing Slots & 1 & 4 & 1 \\
    Checkpoint Interval (s) & 10 & 90 & 10 \\
    \bottomrule
\end{tabular}
\label{tbl:config_parameters}
\end{table}
\setlength{\textfloatsep}{0.1cm}

Profiling runs incorporated a 2-minute stabilization period and a 1-minute latency measurement window. 
We found that the best possible recovery times with \( C_{\text{max}} \) generally span between 90 and 120 seconds, even in scenarios of low workloads. 
This insight led us to set the recovery time constraint (\(RC\)) at 180s for our experiments, establishing the upper limit for acceptable recovery times. 
To accommodate for potential deviations, we defined a maximum timeout of 360s. 
For the optimization phase, the system was configured to perform evaluations every 10 minutes, with a 10-minute time horizon set for the TSF model.
To predict future workloads, we employ online ARIMA for our TSF model using the pmdarima\footnote{\url{https://pypi.org/project/pmdarima/}, Accessed: March 2024} python library.
ARIMA was selected for its proven accuracy and efficiency in forecasting streaming workloads, as detailed in our previous research~\cite{Gontarska2021EvaluationOL}, and is favored due to its low computational demand and minimal data requirements.

All of these default settings are designed to cover a wide range of execution scenarios, providing good performance and should suffice for the majority of users without requiring further adjustments.

\subsection{Baselines Setup}

For our comparative analysis, we included a static configuration alongside two state-of-the-art baselines, both of which are interoperable with Apache Flink. 
As these methods do not focus on optimizing CPU an memory allocations, we assign a full CPU core and 4096 MB of memory to all taskmanagers. 
Moreover, all baseline methods consistently used a 10s checkpoint interval.

\subsubsection{Static Configuration}

In our experiments, we employed a static configuration with 24 worker nodes which aligns with Demeter's \( C_{\text{max}} \) configuration.
This representing the maximum available resources, guarantees the best latencies and recovery times due to the highest resource allocation. 
This baseline serves as the standard for comparing all other methods.

\subsubsection{Flink Reactive}

The first method we evaluated was Apache Flink's reactive mode scheduler~\cite{flink2023reactive}. 
This scheduler dynamically optimizes cluster resources by adjusting to workload variations. 
Monitoring each worker's performance, any deviation from a specified utilization threshold prompts the scheduler to restart the DSP job from its last successful checkpoint with an altered scaleout.
Configured to work with the Kubernetes Horizontal Pod Autoscaler (HPA)\footnote{\url{https://kubernetes.io/docs/tasks/run-application/horizontal-pod-autoscale}, Accessed: March 2024}, our experiments set the HPA to target a CPU utilization of 35\%, aligning with the recommended setup from the reactive mode documentation.
We experimented with several higher utilization targets, but they consistently yielded inferior results.

\subsubsection{DS2 Autoscaler}

The second method we evaluated was the DS2 Autoscaler~\cite{Kalavri2018ThreeSI}, a solution designed for the dynamic scaling of DSP jobs. 
It integrates historical data with real-time metrics, employing forecasting techniques to proactively predict workload variations. 
The system then makes informed scaling decisions, considering both current system state and anticipated workloads. 
This is achieved by optimizing operator parallelism, which adjusts the number of vertices in the execution graph to ensure efficient data processing.
For our evaluation, we used the implementation provided by the Flink Kubernetes Operator, configuring it in such a way as to align it with the other methods.
The system used a stabilization interval of 2 minute and metrics were aggregated over 1-minute windows. 
The target utilization was set at 35\% and a boundary of 15\%. 
Following any scaling adjustments, a 1-minute restart period was observed, and a 5-minute catch-up duration is assumed to guarantee system equilibrium.

\subsection{Experiments}

For comparison, two experiments were conducted using established benchmark jobs and real-world workload simulations.

\subsubsection{Yahoo Streaming Benchmark (YSB) Experiment}

For our first experiment, we used the Yahoo Streaming Benchmark\footnote{\url{https://github.com/yahoo/streaming-benchmarks/}, Accessed: March 2024}. 
This benchmark simulates a streaming advertisement job, structured with multiple advertising campaigns, each containing several individual advertisements. 
Streaming sources retrieve events from a Kafka topic, identify relevant events, and aggregate a windowed count of these events, grouped by campaign. 
A key component of this setup was the deployment of a Redis cluster\footnote{\url{https://redis.io/}, Accessed: March 2024}, which managed campaign and advertisement data, streamlining event generation and data aggregation processes. 
To align the benchmark with our objectives, we enabled checkpointing and replaced the native windowing functionality with the standard Flink implementation. 
We designed a data generator, using a click-through rate dataset\footnote{\url{https://www.kaggle.com/c/avazu-ctr-prediction}, Accessed: March 2024}, to emit events characterized by attributes such as \textit{event\_time}, \textit{event\_type}, and \textit{ad\_id}.
From the initial 10-day dataset, we extracted a 3-day segment and sub-sampled every 4\textsuperscript{th} data point, resulting in a dataset that spans 18 hours.
A graphical representation of the generated workload is provided in~\autoref{fig:evaluation_datasets}(a).
This dataset is characterized by high variability, covering a wide range of processing rates, and lacking a discernible long-term trend.

\subsubsection{Top Speed Windowing (TSW) Experiment}

\begin{figure*}
    \centering
    \subfloat[YSB experiment: Average end-to-end latencies over time.]{
      \includegraphics[width=\columnwidth]{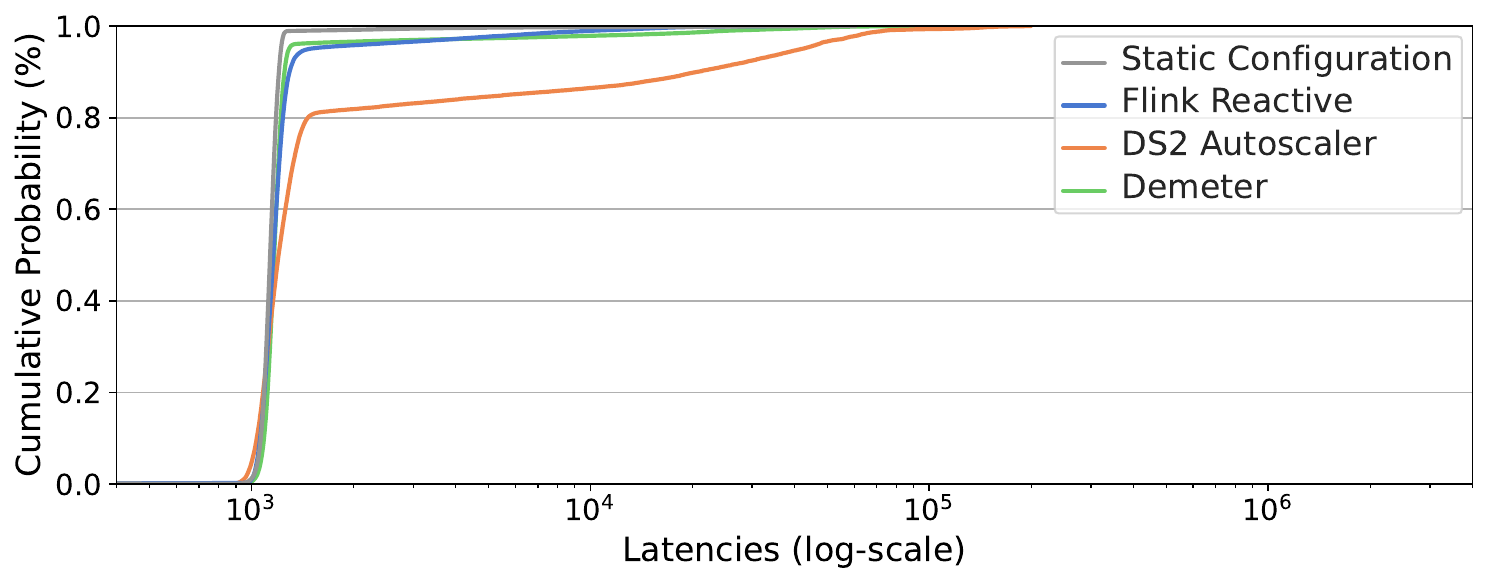}
    }
    \subfloat[TSW experiment: Average end-to-end latencies over time.]{
      \includegraphics[width=\columnwidth]{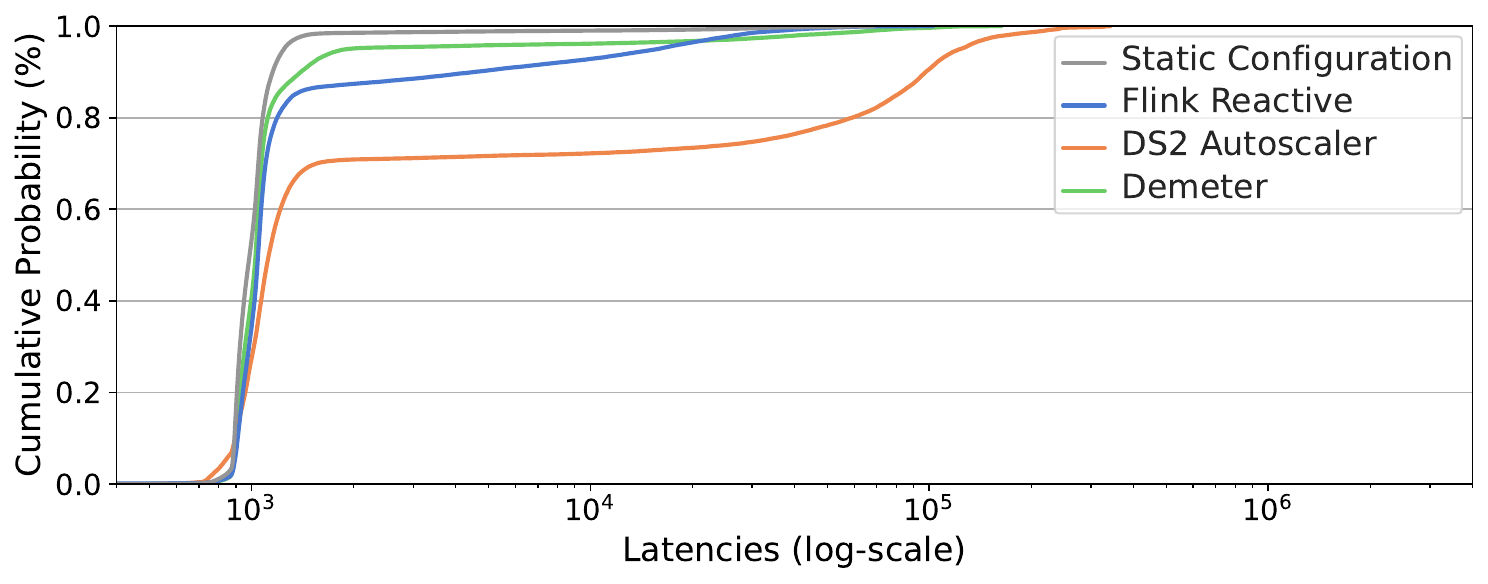}
    }
    \newline
    \subfloat[YSB experiment: Total CPU \& memory usage.]{
      \includegraphics[width=\columnwidth]{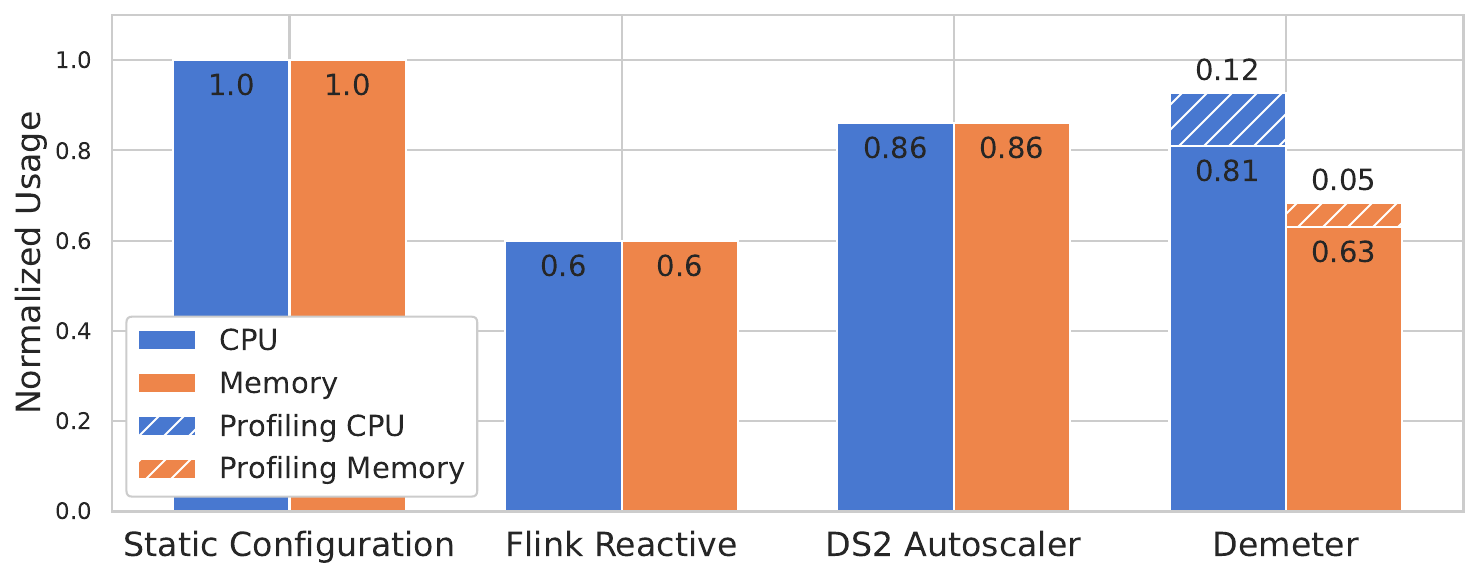}
    }
    \subfloat[TSW experiment: Total CPU \& memory usage.]{
      \includegraphics[width=\columnwidth]{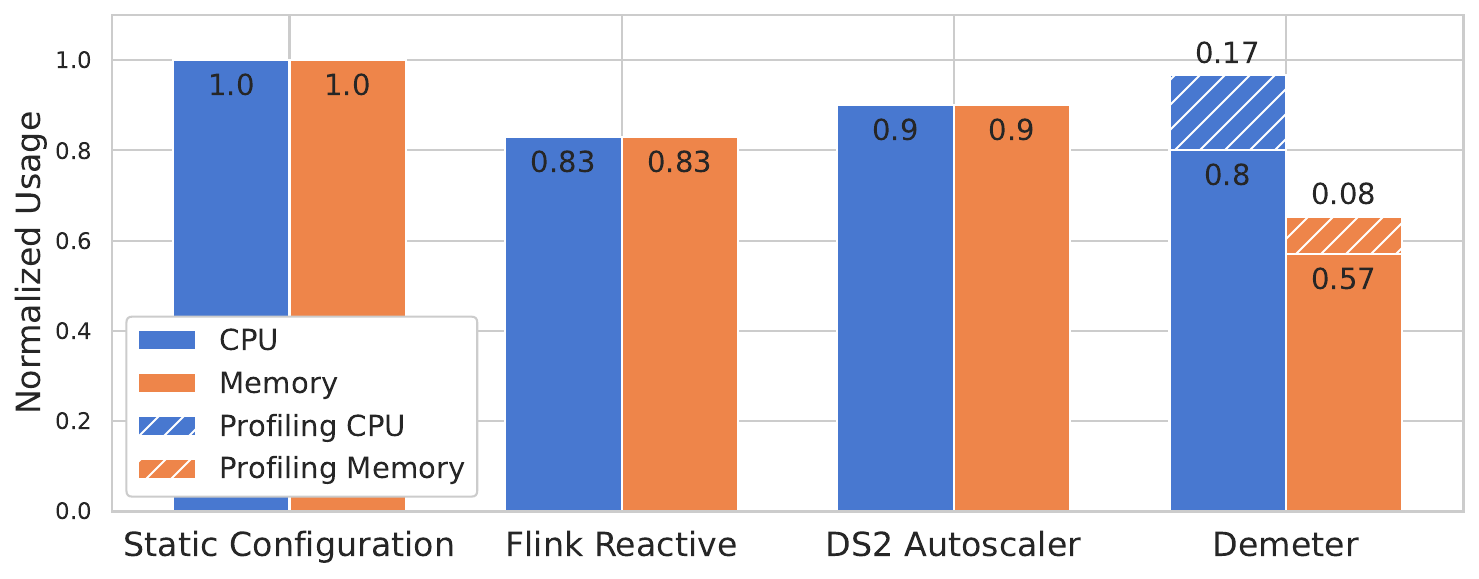}
    }
    \newline
    \subfloat[YSB experiment: Resource usage for largest segment with regression.]{
      \includegraphics[width=\columnwidth]{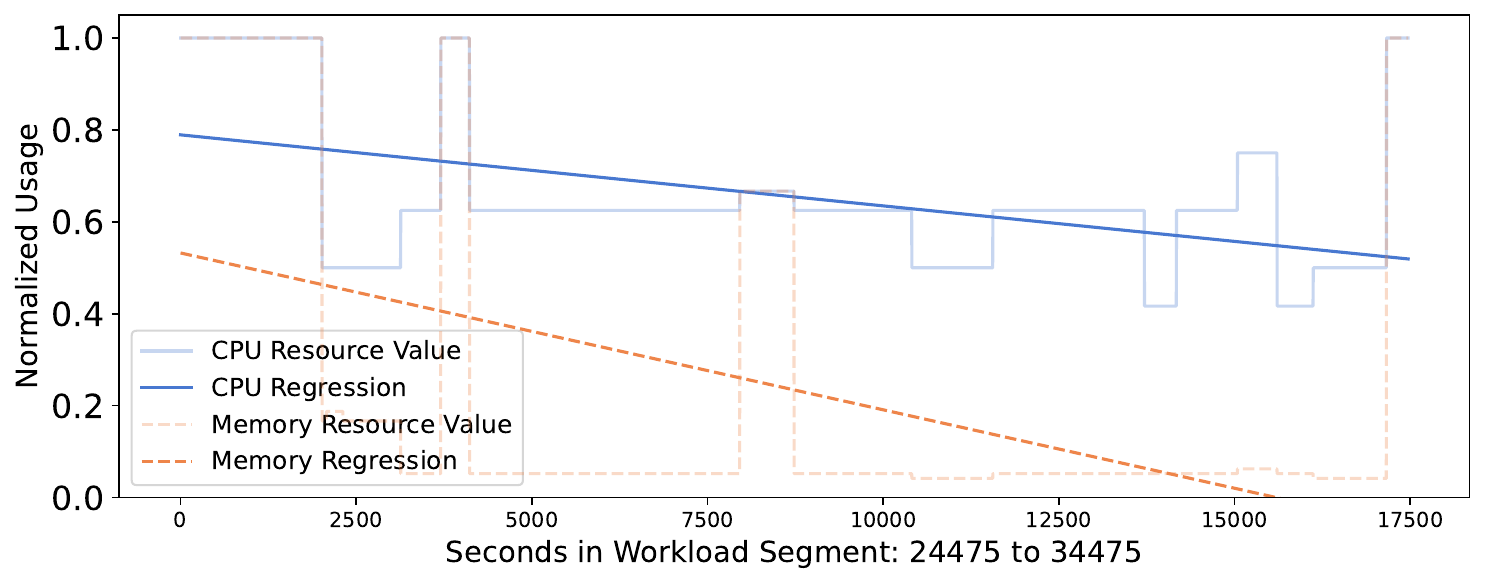}
    }
    \subfloat[TSW experiment: Resource usage for largest segment with regression.]{
      \includegraphics[width=\columnwidth]{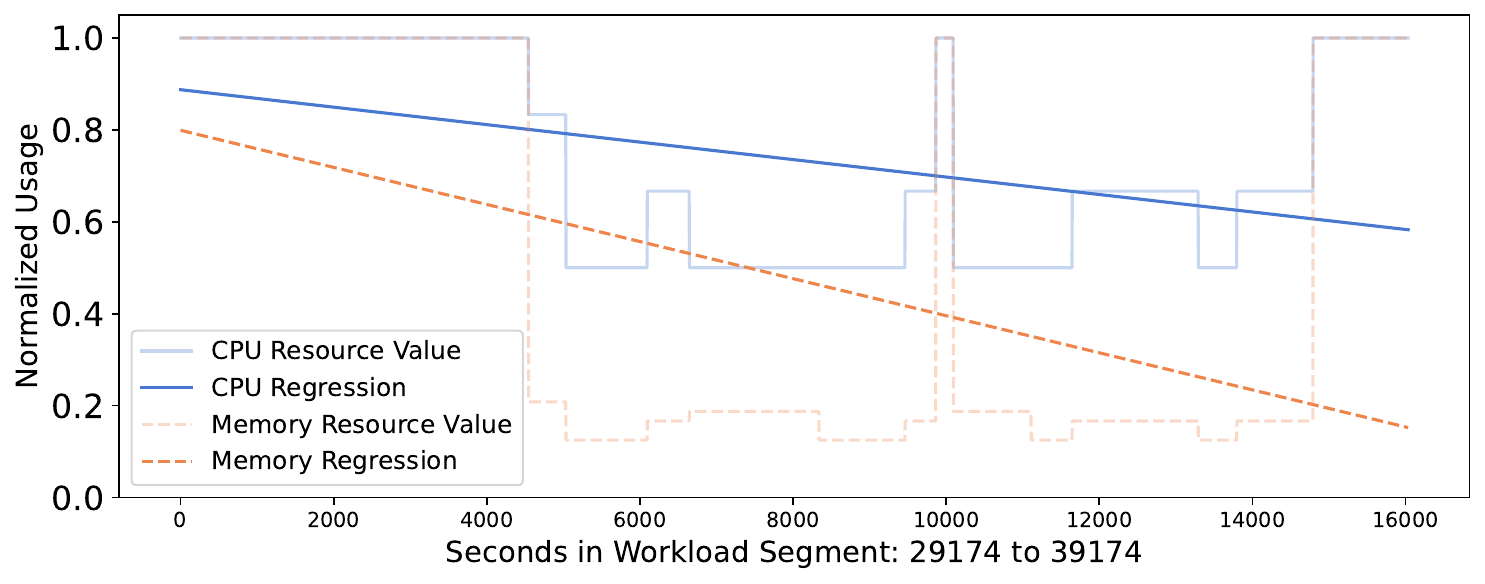}
    }
    \caption{Overview of performance comparison results for Demeter \& state-of-the-art approaches.}
    \label{fig:evaluation_datasets}
\end{figure*}

For our second experiment we used a DSP job derived from the official Flink repository\footnote{\url{https://github.com/apache/flink/}; Accessed: March 2024}. 
The primary focus of the job is on grouped stream windowing, enabling the application of diverse eviction and trigger policies.
Each car-related event consists of attributes such as a unique number plate, the current speed (km/h), the total elapsed distance (meters), and an associated timestamp. 
The goal of the job is to determine the top speed of each car over a span of 50 meters, using only data from the immediate past 10 seconds.
The job was modified to enable it to consume events from and publish results to separate Apache Kafka topics.
To simulate workload variations, represented by the changing number of vehicles over time, we used the Sumo simulation tool to generate a 24-hour workload dataset, specifically employing the TAPASCologne scenario\footnote{\url{https://sumo.dlr.de/docs/}; Accessed: March 2024}. 
Similar to the YSB experiment, we reduced this to 18 hours by sub-sampling every 4\textsuperscript{th} data point and then repeating the resulting workload three times.
This resulted in a dataset characterized by a clear seasonal pattern, with workload rates fluctuating within specific ranges, and a weak upward trend over time.
A graphical representation can be seen in \autoref{fig:evaluation_datasets}(b).
We created a generator program which would produce events constrained by this dataset.

\begin{figure*}
    \centering
    \subfloat[YSB experiment: Workload rates \& all configuration changes.]{
      \includegraphics[width=\textwidth]{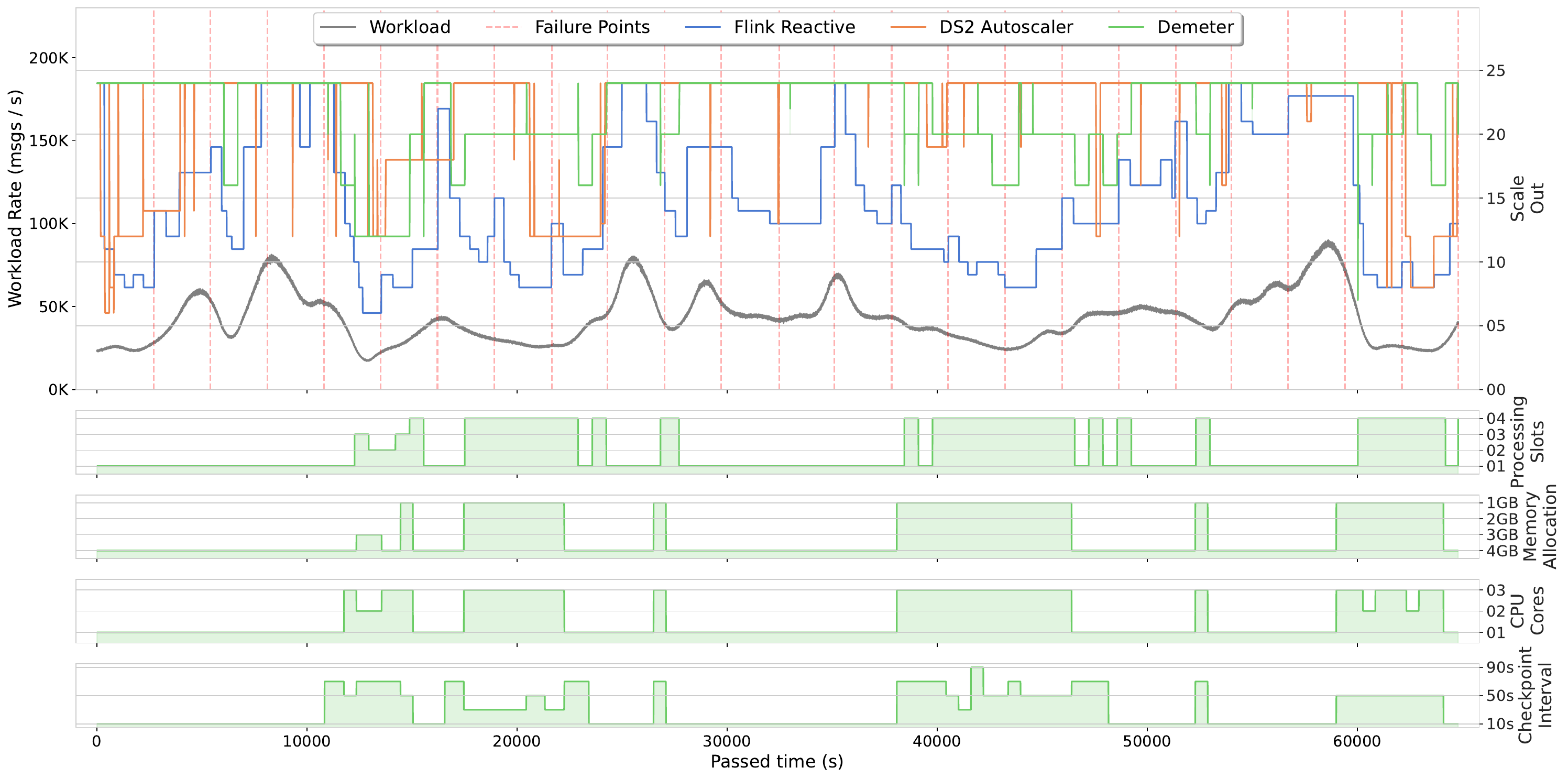}
    }
    \newline
    \subfloat[TSW experiment: Workload rates \& all configuration changes.]{
      \includegraphics[width=\textwidth]{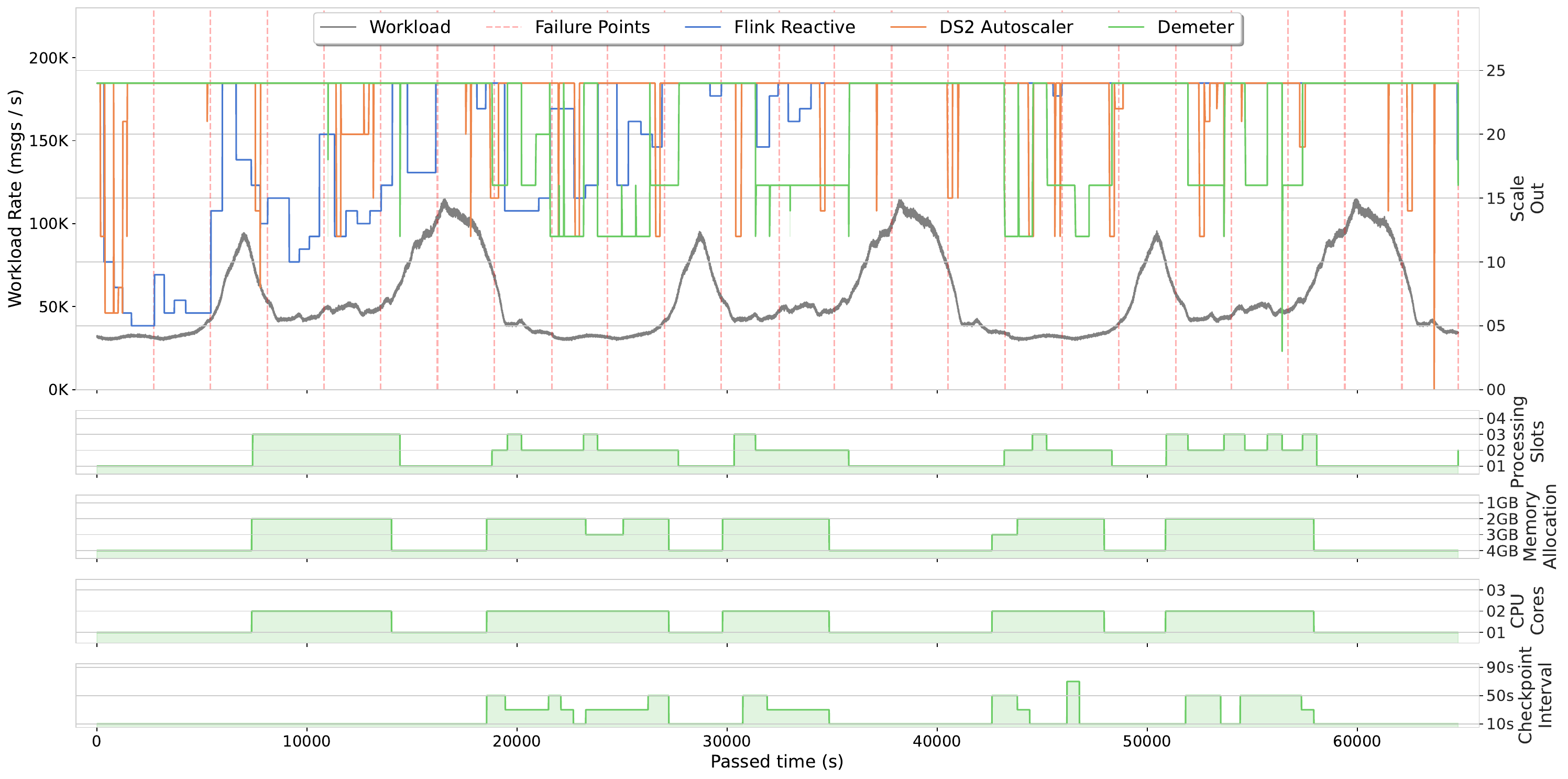}
    }
    \caption{Workloads, failure injections, \& configuration changes for Demeter \& state-of-the-art approaches.}
    \label{fig:reconfigurations}
\end{figure*}

\subsection{Experimental Results}

\begin{table*}[t]
\caption{Recovery times \& number of reconfigurations (\( \Delta \)).}
\centering

\definecolor{hg}{rgb}{0.7, 0.9, 0.7}
\definecolor{hy}{rgb}{1.0, 1.0, 0.5}
\definecolor{hr}{rgb}{1.0, 0.8, 0.8}
\definecolor{hdg}{rgb}{0.85, 0.85, 0.85}

\newcommand{\colorcell}[2]{\setlength{\fboxsep}{1pt}\colorbox{#1}{#2}}

    \subfloat[YSB experiment results.] {
    \small
    \begin{tabular}{p{10mm}|p{3.4mm}p{3.4mm}p{3.4mm}p{3.4mm}p{3.4mm}p{3.4mm}p{3.4mm}p{3.4mm}p{3.4mm}p{3.4mm}p{3.4mm}p{3.4mm}p{3.4mm}p{3.4mm}p{3.4mm}p{3.4mm}p{3.4mm}p{3.4mm}p{3.4mm}p{3.4mm}p{3.4mm}p{3.4mm}p{4mm}|p{4mm}} \toprule
        {} & {$\#1$} & {$\#2$} & {$\#3$} & {$\#4$} & {$\#5$} & {$\#6$} & {$\#7$} & {$\#8$} & {$\#9$} & {$\#10$} & {$\#11$} & {$\#12$} & {$\#13$} & {$\#14$} & {$\#15$} & {$\#16$} & {$\#17$} & {$\#18$} & {$\#19$} & {$\#20$} & {$\#21$} & {$\#22$} & {$\#23$} & {$\Delta$} \\ \midrule
        {Workload} & 30K & 50K & 80K & 55K & 25K & 45K & 30K & 50K & 35K & 50K & 40K & 65K & 45K & 33K & 25K & 35K & 45K & 45K & 50K & 60K & 65K & 25K & 27K & --\\ \midrule[1pt]
        {Static} & \colorcell{hg}{122s} & \colorcell{hg}{95s} & \colorcell{hg}{93s} & \colorcell{hg}{94s} & \colorcell{hg}{95s} & \colorcell{hg}{92s} & \colorcell{hg}{96s} & \colorcell{hg}{96s} & \colorcell{hg}{97s} & \colorcell{hg}{96s} & \colorcell{hg}{95s} & \colorcell{hg}{95s} & \colorcell{hg}{99s} & \colorcell{hg}{94s} & \colorcell{hg}{94s} & \colorcell{hg}{95s} & \colorcell{hg}{95s} & \colorcell{hg}{96s} & \colorcell{hg}{95s} & \colorcell{hg}{94s} & \colorcell{hg}{95s} & \colorcell{hg}{96s} & \colorcell{hg}{95s} & --\\ \midrule
        {Demeter} & \colorcell{hg}{123s} & \colorcell{hg}{97s} & \colorcell{hg}{97s} & \colorcell{hg}{98s} & \colorcell{hg}{90s} & \colorcell{hg}{95s} & \colorcell{hg}{90s} & \colorcell{hg}{89s} & \colorcell{hdg}{NR} & \colorcell{hg}{120s} & \colorcell{hg}{95s} & \colorcell{hg}{96s} & \colorcell{hg}{100s} & \colorcell{hg}{96s} & \colorcell{hdg}{NR} & \colorcell{hg}{90s} & \colorcell{hg}{87s} & \colorcell{hdg}{NR} & \colorcell{hg}{97s} & \colorcell{hg}{96s} & \colorcell{hg}{95s} & \colorcell{hg}{96s} & \colorcell{hg}{140s} & 33 \\ \midrule
        {Reactive} & \colorcell{hy}{201s} & \colorcell{hy}{200s} & \colorcell{hg}{95s} & \colorcell{hg}{145s} & \colorcell{hy}{182s} & \colorcell{hy}{255s} & \colorcell{hg}{146s} & \colorcell{hy}{252s} & \colorcell{hg}{95s} & \colorcell{hg}{165s} & \colorcell{hy}{190s} & \colorcell{hdg}{NR} & \colorcell{hg}{143s} & \colorcell{hg}{136s} & \colorcell{hy}{193s} & \colorcell{hg}{164s} & \colorcell{hy}{185s} & \colorcell{hy}{190s} & \colorcell{hy}{266s} & \colorcell{hg}{135s} & \colorcell{hg}{152s} & \colorcell{hdg}{NR} & \colorcell{hy}{205s} & 87 \\ \midrule
        {DS2} & \colorcell{hy}{187s} & \colorcell{hg}{95s} & \colorcell{hg}{125s} & \colorcell{hy}{204s} & \colorcell{hy}{330s} & \colorcell{hg}{122s} & \colorcell{hr}{6m+} & \colorcell{hg}{125s} & \colorcell{hr}{6m+} & \colorcell{hg}{126s} & \colorcell{hg}{95s} & \colorcell{hr}{6m+} & \colorcell{hg}{125s} & \colorcell{hg}{129s} & \colorcell{hdg}{NR} & \colorcell{hg}{95s} & \colorcell{hg}{91s} & \colorcell{hg}{120s} & \colorcell{hr}{6m+} & \colorcell{hdg}{NR} & \colorcell{hg}{155s} & \colorcell{hg}{95s} & \colorcell{hdg}{NR} & 77 \\ \midrule\bottomrule[1pt]
    \end{tabular}
    }
    \newline
    \subfloat[TSW experiment results.] {
    \small
    \begin{tabular}{p{10mm}|p{3.4mm}p{3.4mm}p{3.4mm}p{3.4mm}p{3.4mm}p{3.4mm}p{3.4mm}p{3.4mm}p{3.4mm}p{3.4mm}p{3.4mm}p{3.4mm}p{3.4mm}p{3.4mm}p{3.4mm}p{3.4mm}p{3.4mm}p{3.4mm}p{3.4mm}p{3.4mm}p{3.4mm}p{3.4mm}p{4mm}|p{4mm}} \toprule
        {} & {$\#1$} & {$\#2$} & {$\#3$} & {$\#4$} & {$\#5$} & {$\#6$} & {$\#7$} & {$\#8$} & {$\#9$} & {$\#10$} & {$\#11$} & {$\#12$} & {$\#13$} & {$\#14$} & {$\#15$} & {$\#16$} & {$\#17$} & {$\#18$} & {$\#19$} & {$\#20$} & {$\#21$} & {$\#22$} & {$\#23$} & {$\Delta$} \\ \midrule
        {Workload} & 30K & 50K & 80K & 55K & 25K & 45K & 30K & 50K & 35K & 50K & 40K & 65K & 45K & 33K & 25K & 35K & 45K & 45K & 50K & 60K & 65K & 25K & 27K & --\\ \midrule[1pt]
        {Static} & \colorcell{hg}{95s} & \colorcell{hg}{95s} & \colorcell{hg}{95s} & \colorcell{hg}{94s} & \colorcell{hg}{96s} & \colorcell{hg}{155s} & \colorcell{hg}{98s} & \colorcell{hg}{96s} & \colorcell{hg}{100s} & \colorcell{hg}{97s} & \colorcell{hg}{96s} & \colorcell{hg}{152s} & \colorcell{hg}{127s} & \colorcell{hg}{126s} & \colorcell{hg}{95s} & \colorcell{hg}{106s} & \colorcell{hg}{96s} & \colorcell{hg}{96s} & \colorcell{hg}{98s} & \colorcell{hg}{93s} & \colorcell{hg}{97s} & \colorcell{hg}{166s} & \colorcell{hg}{95s} & --\\ \midrule
        {Demeter} & \colorcell{hg}{96s} & \colorcell{hg}{95s} & \colorcell{hg}{90s} & \colorcell{hg}{90s} & \colorcell{hg}{120s} & \colorcell{hg}{127s} & \colorcell{hg}{150s} & \colorcell{hdg}{NR} & \colorcell{hg}{90s} & \colorcell{hg}{120s} & \colorcell{hg}{100s} & \colorcell{hg}{90s} & \colorcell{hg}{116s} & \colorcell{hg}{160s} & \colorcell{hg}{95s} & \colorcell{hdg}{NR} & \colorcell{hg}{89s} & \colorcell{hg}{95s} & \colorcell{hg}{123s} & \colorcell{hg}{120s} & \colorcell{hg}{120s} & \colorcell{hg}{160s} & \colorcell{hg}{120s} & 30 \\ \midrule
        {Reactive} & \colorcell{hg}{158s} & \colorcell{hy}{210s} & \colorcell{hg}{171s} & \colorcell{hg}{175s} & \colorcell{hy}{205s} & \colorcell{hy}{200s} & \colorcell{hy}{199s} & \colorcell{hg}{170s} & \colorcell{hg}{178s} & \colorcell{hg}{95s} & \colorcell{hg}{146s} & \colorcell{hy}{198s} & \colorcell{hy}{225s} & \colorcell{hg}{150s} & \colorcell{hy}{185s} & \colorcell{hy}{220s} & \colorcell{hg}{170s} & \colorcell{hy}{192s} & \colorcell{hg}{115s} & \colorcell{hg}{170s} & \colorcell{hg}{175s} & \colorcell{hg}{146s} & \colorcell{hg}{128s} & 49 \\ \midrule
        {DS2} & \colorcell{hg}{125s} & \colorcell{hg}{95s} & \colorcell{hg}{95s} & \colorcell{hg}{94s} & \colorcell{hg}{125s} & \colorcell{hy}{275s} & \colorcell{hr}{6m+} & \colorcell{hy}{226s} & \colorcell{hg}{95s} & \colorcell{hr}{6m+} & \colorcell{hg}{127s} & \colorcell{hg}{124s} & \colorcell{hg}{125s} & \colorcell{hg}{157s} & \colorcell{hdg}{NR} & \colorcell{hg}{135s} & \colorcell{hdg}{NR} & \colorcell{hr}{6m+} & \colorcell{hg}{95s} & \colorcell{hg}{126s} & \colorcell{hg}{95s} & \colorcell{hr}{6m+} & \colorcell{hg}{97s} & 73 \\ \midrule\bottomrule[1pt]
    \end{tabular}
    }
    \newline
    \label{tbl:recovery_times}
\end{table*}
\setlength{\textfloatsep}{0.1cm}

After completing the experiments, we analyzed the performance metrics, with Figures \ref{fig:reconfigurations}(a) and \ref{fig:reconfigurations}(b) showing the workload rates over time and scaleout decisions for all methods. 
The data shows that Demeter tends to favor higher scale-out values, aligning with expectations as newly encountered workload rates trigger the use of $C_{max}$, and the safety buffer hyper-parameter increases the optimal amount of resources by 30\%. 
Moreover, an increase in CPU usage was consistently observed in the TSW experiment across all executions, including static configurations. 
Further analysis identified a statistically significant weak upward trend within this dataset, likely explaining the observed increase in CPU usage.

\subsubsection{Average End-to-End Latency Results:}

Our analysis first examines the $L_{avg}$ of the static baseline, Demeter, and comparative methods, shown in Figures \ref{fig:evaluation_datasets}(a) and \ref{fig:evaluation_datasets}(b) using an empirical cumulative distribution function. 
The static configuration consistently had latencies near 1000ms, a high proportion of optimal latencies. 
Among the optimization methods, Demeter led with the most latencies in the optimal range, followed by Flink Reactive and then DS2. 
In the YSB experiment, Demeter and Flink Reactive achieved near-optimal latencies in about 95\% of cases, compared to DS2's 80\%. 
In the TSW experiment, Demeter maintained over 95\% in optimal latencies, with Flink Reactive at about 85\% and DS2 at 70\%.

\subsubsection{Recovery time Results:}

Recovery times were manually measured by analyzing consumer lag and throughput rate metrics (refer to Section \ref{sec:approach_profiling}). 
The results, detailed in Table\autoref{tbl:recovery_times}, use color highlights to indicate performance: green for recovery times under the 180s recovery time constraint, yellow for exceeding this constraint, and red for surpassing the 6-minute maximum.
The static configuration, with maximum resources, set a benchmark with the fastest recovery times, averaging 96s in the YSB experiment. 
Demeter showed a minimal deviation in recovery times of 3.21\% compared to this benchmark, while Reactive and DS2 had deviations of 82.79\% and 77.59\%, respectively.
In the TSW experiment, the static configuration's average recovery time was 107s, with Demeter at a 5.17\% deviation, maintaining consistent performance. 
Reactive and DS2 had higher deviations of 62.78\% and 51.62\%.
These findings underscore Demeter's ability to closely match the optimal recovery times of the static configuration, unlike the larger variances seen with other methods. 
Additionally, Demeter was the method which initiated the least number of reconfigurations ($\Delta$).
'No Result' (NR) entries in our table reflect the dynamic nature of jobs; reconfigurations for exactly-once processing sometimes overlap with failure injections, leading to unsuccessful recovery attempts.

\subsubsection{Resource Usage Results:}

Figures \ref{fig:evaluation_datasets}(c) and \ref{fig:evaluation_datasets}(d) display the cumulative CPU and memory usage for both experiments, normalized against the maximum resource usage benchmarked at 100\%. 
For Demeter, both the target job's resource usage and the cost of profiling are included. 
As expected, the static configuration consistently showed the highest consumption. 
In the YSB experiment, Flink Reactive achieved a 40\% reduction in CPU and memory usage relative to the static configuration. 
Demeter demonstrated a 19\% reduction in CPU usage and 37\% in memory, while DS2 recorded a 14\% improvement.
Including Demeter's profiling costs, the net resource savings were 7\% in CPU and 32\% in memory usage compared to the static configuration. 
Flink reactive was the best performer for this experiment.
In the TSW experiment, Demeter achieved a 20\% reduction in CPU usage and 43\% in memory compared to the static configuration. 
Flink Reactive showed a 17\% reduction, and DS2 had a 10\% improvement. 
After factoring in profiling costs, Demeter's overall efficiency was a 3\% saving in CPU and 35\% in memory. 
In this case, Demeter had the most efficient memory usage, while Flink Reactive was the most efficient in terms of CPU usage.

\subsection{Experimental Discussion}

In both experiments, Demeter maintained latencies and recovery times close to those of the static configuration, while also enhancing resource efficiency.
Flink Reactive showed good performance in terms of latencies in the YSB experiment; however, it was not able to achieve comparable results in the TSW experiment.
Moreover, both Flink Reactive and DS2 had variable recovery times, often substantially longer than the static configuration, indicating inconsistent performance.
For methods aimed at enhancing resource efficiency to be considered effective, maintaining a high Quality of Service is fundamental.
If the service is not reliably available, the resulting benefits of resource savings are irrelevant.
This is further evidenced by Demeter initiating significantly fewer reconfigurations across both experiments, thereby minimizing disruptions to the service.
In the experiments, the only two methods that consistently met this requirement were the static configuration and Demeter.

While the initial analysis indicates Demeter's modest CPU usage improvement of 7\% and 3\%, and a substantial decrease in memory usage by 32\% and 35\% compared to the static configuration, continuing the experiment could reveal further enhancements. 
Extending execution beyond 18 hours would enable more profiling and the selection of increasingly optimized configurations, thus reducing the time spent at $C_{max}$. 
Assuming similar workload rates, we anticipate a further decrease in resource usage over time.
This trend is illustrated in Figures \ref{fig:evaluation_datasets}(e) and \ref{fig:evaluation_datasets}(f), which show CPU and memory usage over time, concentrating on the workload segments with the highest data points in each experiment. 
We included regression lines to highlight the resource usage trends. 
The data shows a general decrease in resource utilization, with CPU usage potentially dropping by 20\% to 30\%, and memory usage by 50\% to 70\%.

Similarly, if the experiment were to be extended, the annealing factor applied to profiling would lead to a reduction in profiling executions as knowledge is accumulated.
Consequently, the cost of profiling is expected to progressively decrease, further enhancing Demeter's resource efficiency. 
Considering this, it is possible that Demeter would not only maintain its high performance but also improve upon the resource usage results from the other methods.

%YSB: 11.7\% CPU, 5.4\% memory; 92.7 68.4
%TSW: 16.6\% CPU, 8.2\% memory; 96.6 65.2

%ADS Experiment

%static avg: 96,260869565
%demeter: 99,35
%Reactive: 175,952380952
%DS2: 170,95

%percentage error:
%demeter: 3,209123758
%Reactive: 82,787026283
%DS2: 77,590334237

%CARS Experiment

%static: 107,130434783
%demeter: 112,666666667
%Reactive: 174,391304348
%DS2: 162,428571429

%percentage error:
%demeter: 5,167748918
%Reactive: 62,784090909
%DS2: 51,61757885

\section{Related Work}

In this section, we explore the work related to our own, examining the domain of automatic configuration optimization. 

\subsection{Batch Processing Optimization}

In batch processing, configuring resources to meet runtime targets is a key challenge. 
In exploring methods similar to our own, we examined Bayesian Optimization in CherryPick~\cite{Alipourfard2017CherryPickAU}, which efficiently identifies near-optimal configurations for batch jobs using adequately precise performance models.
Karasu~\cite{ScheinertWWTWK2023} likewise makes use of this, employing MOBO and RGPE modeling for resource optimization and introducing a strategy for profiling with shared user data to overcome the cold start problem.
Other methods focus on modeling batch job scaleout behaviors, using historical data for resource allocation—similar to our data gathering through profiling runs. 
Ernest~\cite{Venkataraman2016ErnestEP} predicts cloud resource needs by running jobs with subsets of inputs and different configurations.
Bell~\cite{Thamsen2016SelectingRF} uses existing workload data from recurring jobs for its predictions, eliminating isolated training runs. 
In our previous work, we presented Enel~\cite{ScheinertZTGWAK21} which adopts a context-aware, graph-based approach for more detailed insights within batch jobs' synchronization barriers by incorporating runtime metrics.
Taking inspiration from these methods, Demeter applies this to the stream processing domain.

\subsection{Stream Processing Optimization}

In stream processing, optimization approaches primarily focus on elastic scaling of resources to adapt to runtime conditions. 
A number of these focus on threshold monitoring and re-configuring after a predefined threshold has been violated.
Gedik et al. in \cite{GSH+14} introduce a control algorithm for IBM Infosphere Streams~\cite{Biem2010IBMIS} that reacts to congestion, while Dhalion~\cite{FAG+17} applies policy-driven strategies for Heron~\cite{Kulkarni2015TwitterHS}. 
TWRES~\cite{HuKZ19} uses TSF for predicting future workloads and adjusts resources based on a latency constraint.
Prompt~\cite{Abdelhamid2020PromptDD} is a data partitioning scheme specifically for micro-batch DSPs, focused on maintaining latency guarantees through a threshold-based elasticity technique that dynamically adjusts execution parallelism.
Apache Flink’s Reactive Mode~\cite{flink2023reactive} automates the scaling process, dynamically adjusting resources to workload variations. 
Elastic Spark Streaming~\cite{databricks2023elasticSpark} also adopts a similar automated scaling approach, enhancing resource management in Spark environments.

However, the reliability of threshold-based autoscalers can be affected by the transient nature of shared computing environments and DSP job behaviors, leading to unnecessary adjustments.
As a result, stream processing research has more recently adopted data-driven approaches using performance modeling for scaling decisions. 
Petrov et al.\cite{PETROV2018109} detail a model that bases scaling actions on latency measurements, and DS2\cite{Kalavri2018ThreeSI} uses historical and real-time data for workload forecasting to dynamically scale streaming dataflows. 
In our previous work with Phoebe~\cite{Geldenhuys2022PhoebeQD}, initial profiling was conducted to establish models that map scaleout and workload rates to latency and recovery times. 
TSF was then employed to predict future workloads, allowing for dynamic rescaling of resources aimed at maintaining stable latencies and achieving optimal recovery times.
However, these data-driven methods can be limited by the availability of historical data, challenging their accuracy in dynamic environments.
In contrast, Demeter concurrently addresses multiple configuration parameters and incorporates an efficient profiling method for data gathering, enabling more comprehensive optimization in stream processing.

In the context of Bayesian Optimization, a number of methods have been proposed for the optimization of configuration parameters in DSP systems.
Fischer et al., in \cite{Fischer2015MachinesTM}, suggest a technique for tuning multiple parameters that, while effective for Apache Storm---an older DSP system---is not widely applicable, hence not offering a generalizable solution.
In \cite{Jamshidi2016AnUA}, Jamshidi et al. present a method for tuning configuration parameters to reduce latency and maintain throughput. 
Their results are positive, but the paper does not discuss workload rates, leading to an implicit assumption that their optimization is designed for static workloads.
ContTune~\cite{Lian2023ContTuneCT} focuses on a single configuration parameter which employs Conservative Bayesian Optimization to fine-tune DSP job parallelism while ensuring SLA adherence. 
However, these methods overlook fault tolerance and SLA-specific recovery time considerations.

\subsection{Checkpoint Optimization}

Another area of related work is fault tolerance, particularly the optimization of the checkpoint and rollback recovery mechanism in DSP systems. 
Here the system's state is periodically saved, allowing for restarts from the latest checkpoint after a failure. 
Our approach, which involves optimizing the checkpoint interval to enhance system performance, is related to other methods that adjust this parameter. 
In high-performance computing, some methods determine the mean time to failure of cluster nodes and modify the checkpoint interval to reduce downtime from failures~\cite{Y74,D03,D06}. 
In the specific context of stream processing, ~\cite{JHK20} explore the effects of system failures and configurations on recovery, aiding in the development of more efficient checkpoint scheduling strategies.
Our work differs from theirs in that while they do not seek optimal configurations for DSP jobs at runtime, Demeter actively does.
In our previous work, we also investigated parameter auto-tuning of DSP jobs to improve end-to-end latencies and recovery time, yet focused on optimizing checkpoint intervals while assuming scaleouts to be static~\cite{Geldenhuys2019EffectivelyTS, Geldenhuys2020ChironOF}.

\section{Conclusion}

In this paper, we presented Demeter, a method designed to enhance the resource efficiency of DSP jobs running in dynamically changing environments. 
By utilizing TSF to predict future workloads and MOBO to model runtime behaviors, Demeter effectively decides itself when to initiate short-lived parallel profiling runs and when to proceed with optimization. 
This approach guides the adjustment of multiple configuration parameters, providing near-optimal performance as workload rates vary. 
Our evaluation results show that Demeter not only matches the high performance of over-provisioned static configurations in terms of average end-to-end latencies and recovery times but also significantly improves resource efficiency. 
Specifically, Demeter showed improvements of 7\% and 3\% in CPU usage, and more substantially, 32\% and 35\% in memory usage.
As the experiments show, with further execution, resource efficiency would increase, leading to additional cost savings, which is particularly important in cloud environments where memory resources are more expensive than CPU resources.
Furthermore, Demeter is designed for scenarios requiring strict data consistency and the need for exactly-once processing guarantees.

\section*{Acknowledgment}
For the purpose of open access, the authors have applied a Creative Commons Attribution (CC BY) license to any Author Accepted Manuscript version arising from this submission.

\balance
\bibliographystyle{ACM-Reference-Format}
\bibliography{references} 

\end{document}